\documentclass[
    amsmath,
    amssymb,
    superscriptaddress,
    aps,
    prx,
    reprint,
    floatfix,
]{revtex4-2}

\usepackage{silence}  
\usepackage[export]{adjustbox}
\usepackage[dvipsnames]{xcolor}
\usepackage{amsmath}
\usepackage{amssymb}
\usepackage{booktabs}
\usepackage{comment}
\usepackage{glossaries}
\usepackage{graphicx}
\usepackage[version=4]{mhchem}
\usepackage{siunitx}
\usepackage{tabularx}
\usepackage[colorlinks=true]{hyperref}
\hypersetup{
    pdffitwindow=false,
    pdfstartview={FitH},
    pdfnewwindow=true,
    colorlinks=true,
    linkcolor=blue,
    citecolor=blue,
    filecolor=blue,
    urlcolor=blue,
}

\hypersetup{pdfauthor={Jakub Fojt, Tuomas P. Rossi, Paul Erhart}}
\hypersetup{pdftitle={rhodent: A Python package for analyzing real-time TDDFT response}}

\setacronymstyle{long-short}
\newacronym{dft}{DFT}{density functional theory}
\newacronym{dos}{DOS}{density of states}
\newacronym{fwhm}{FWHM}{full width at half-maximum}\newacronym{gllbsc}{GLLB-sc}{Gritsenko-van Leeuwen-van Lenthe-Baerends-solid-correlation}
\newacronym{hc}{HC}{hot carrier}
\newacronym{he}{HE}{hot electron}
\newacronym{hh}{HH}{hot hole}
\newacronym{hxc}{Hxc}{Hartree-exchange-correlation}
\newacronym{homo}{HOMO}{highest occupied molecular orbital}
\newacronym{ks}{KS}{Kohn-Sham}
\newacronym{lspr}{LSPR}{localized surface plasmon resonance}
\newacronym{lcao}{LCAO}{linear combination of atomic orbitals}
\newacronym{lumo}{LUMO}{lowest unoccupied molecular orbital}
\newacronym{lp}{LP}{lower polariton}
\newacronym{np}{NP}{nanoparticle}
\newacronym{paw}{PAW}{projector augmented wave}
\newacronym{pdos}{PDOS}{projected density of states}
\newacronym{rttddft}{rt-TDDFT}{real-time time-dependent density functional theory}
\newacronym{tcm}{TCM}{transition contribution map}
\newacronym{up}{UP}{upper polariton}
\newacronym{xc}{XC}{exchange-correlation}

\makeatletter
\let\oldtheequation\theequation
\renewcommand\tagform@[1]{\maketag@@@{\ignorespaces#1\unskip\@@italiccorr}}
\renewcommand\theequation{(\oldtheequation)}
\makeatother

\DeclareSIUnit\angstrom{\text{\AA}}
\DeclareSIUnit\Da{\text{Da}}

\renewcommand{\vec}[1]{\boldsymbol{#1}}
\newcommand{\rhodent}{\textsc{rhodent}}

\newcommand{\chalmers}{
    Department of Physics,
    Chalmers University of Technology,
    SE 412~96 Gothenburg, Sweden
}

\newcommand{\cscaddress}{
    CSC -- IT Center for Science Ltd.,
    P.O. Box 405, 02101 Espoo, Finland
}

\begin{document}

\title{rhodent: A Python package for analyzing real-time TDDFT response}

\author{Jakub Fojt}
\affiliation{\chalmers}
\author{Tuomas P. Rossi}
\affiliation{\cscaddress}
\author{Paul Erhart}
\email{erhart@chalmers.se}
\affiliation{\chalmers}

\begin{abstract}
\Glsentryfull{rttddft} is a well-established method for studying the dynamic response of matter in the femtosecond or optical range.
In this method, the \glsentryfull{ks} wave functions are propagated forward in time, and in principle, one can extract any observable at any given time.
Alternatively, by taking a Fourier transform, spectroscopic quantities can be extracted.
There are many publicly available codes implementing \glsentryshort{rttddft}, which differ in their numeric solution of the \glsentryshort{ks} equations, their available \glsentrylong{xc} functionals, and in their analysis capabilities.
For users of \glsentryshort{rttddft}, this is an inconvenient situation because they may need to use a numerical method that is available in one code, but an analysis method available in another.
Here, we introduce \rhodent{}, a modular Python package for processing the output of \glsentryshort{rttddft} calculations.
Our package can be used to calculate hot-carrier distributions, energies, induced densities, and dipole moments, and various decompositions thereof.
In its current version, \rhodent{} handles calculation results from the \textsc{gpaw} code, but can readily be extended to support other \glsentryshort{rttddft} codes.
Additionally, under the assumption of linear response, \rhodent{} can be used to calculate the response to a narrow-band laser, from the response to a broad-band perturbation, greatly speeding up the analysis of frequency-dependent excitations.
We demonstrate the capabilities of \rhodent{} via a set of examples, for systems consisting of Al and Ag clusters and organic molecules.
\end{abstract}

\maketitle

\section{Introduction}
\Gls{rttddft} is a well-established method for studying the response of electronic systems \cite{ProIsb16, LiGovIsb20}.
It has been used for spectroscopy calculations \cite{YabBer96, KuiSakRos15, SanSanOvi11, ZhaFeiRub14, SinGarWei17, MatLub18, SenLinKud19, RosSheErh19},
including circular dichroism \cite{LeeYabBer11, MakRosLar21}, X-rays \cite{HerZhuAla23}, non-linear spectroscopy \cite{TakVilReh07}, and high-harmonic generation \cite{CasMarAlo04, ChuGro12, DinVanEic13}, as well as to study electron \cite{MenKax10, FalRozBri14, LonPre14, LiaGuaHu18, KumRosKui19, FojRosKui22} and spin dynamics \cite{LiPuGao24} on femtosecond timescales.
In essence, one simply takes the ground state calculated with \gls{ks}-\gls{dft} \cite{HohKoh64, KohSha65} and numerically propagates the single-particle Schrödinger equation forward in time, under the influence of something driving the system out of the ground state, typically an external electric field.
There are various codes implementing \gls{rttddft}, including \textsc{gpaw} \cite{MorLarKui24}, \textsc{octopus} \cite{TanOliAnd20, CasMarRub04}, \textsc{turbomole} \cite{MulShaSie20}, Siesta \cite{GarPapAkh20}, and \textsc{cp2k} \cite{HutIanSch14}.
These codes differ in implementation details and numerical representations of the \gls{ks} wave functions.

There is, however, a need for modular libraries \cite{Leh23}.
It may be desirable for some users to employ one code, due to its numerical implementation details, while using the analysis capabilities of another.
Without modular libraries, the only solution to this problem is for the developer to either port the analysis capabilities or the numerical implementation to the other code.
Here, we introduce \rhodent{} -- a Python package for processing the response from \gls{rttddft} calculations.
In its current version, calculation outputs from \textsc{gpaw} are supported, but the modular design of \rhodent{} prepares it for future integration with other programs.
\rhodent{} is open source and comes with proper unit- and integration tests, easing development, even for external contributors.
An extensive documentation is available online \cite{rhodent}.

\section{Program overview}

The \rhodent{} package is written in Python and designed in a modular and object-oriented fashion.
The key components are the \texttt{Response} and \texttt{Calculator} objects.
The \texttt{Response} class allows reading output files generated in a previous \gls{rttddft} calculation (outside \rhodent{}), and transforms the response into the appropriate form for \rhodent{}, either in the time or frequency domain.
There are several implemented \texttt{Response} classes, for use with different \gls{rttddft} output files.
The \texttt{Calculator} is responsible for computing the observables at various times in the simulation, or the Fourier transform thereof.
Several \texttt{Calculator} classes are available.

The modular structure of \rhodent{} opens up the possibility to, with relatively little effort, extend the functionality of the code.
In order to interface \rhodent{} to another \gls{rttddft} code than \textsc{gpaw}, which is currently supported, a new \texttt{Response} class needs to be added.
Similarly, in order to compute a new type of observable, one in principle only needs to implement a new \texttt{Calculator}.

\section{Implementation and theory}

The central quantity that \rhodent{} works with is the time-dependent KS density matrix in the basis of ground-state \gls{ks} orbitals
\begin{equation}
    \begin{split}
    &\rho_{nn'}(t) = \\
    & \sum_k f_k
       \int \psi^*_{n'}(\vec{r}, 0) \psi_k(\vec{r}, t) \mathrm{d}{\vec{r}}
       \int \psi^*_k(\vec{r}', t) \psi_{n}(\vec{r'}, 0) \mathrm{d}{\vec{r}'},
    \end{split}
    \label{eq:resp:rho_nn}
\end{equation}
where $\psi_k(\vec{r}, t)$ are the time-dependent \gls{ks} wave functions and $f_k$ their occupation numbers (including a spin degeneracy of 2).
In particular, we are interested in the induced density matrix $\delta\rho_{nn'}(t) = \rho_{nn'}(t) - \rho_{nn'}(0)$.
This quantity contains all information about the system response to some time-dependent perturbation $\hat{v}(t)$, and is obtained from an \gls{rttddft} calculation (e.g., using \textsc{gpaw}).
We note here that, since \rhodent{} currently is implemented for non-periodic and spin-paired systems, the only quantum number of the \gls{ks} orbitals is $n$.
We are restricted to fixed-atom calculations, so the time-dependence comes only from electronic degrees of freedom.

\subsection{Obtaining the response}

In \rhodent{}, we can work with the response in the time domain, which is obtained directly from the \gls{rttddft} calculation, or in the frequency domain.
For the latter, we define the normalized Fourier transform
\begin{align}
    \frac{\delta \rho_{nn'}(\omega)}{v(\omega)} = \frac{\int_{0}^\infty\delta\rho_{nn'}(t)e^{i\omega t}\mathrm{d}t}{\int_{0}^\infty v(t)e^{i\omega t}\mathrm{d}t},
\end{align}
where $v(t)$ is the scalar amplitude of the perturbation, and the lower bound on the integrals can be taken to be zero, because the integrands are zero before this time.
This quantity is related to the Casida eigenvectors and gives similar information as the solution of the Casida equation \cite{RosKuiPus17}.

\subsubsection{In the frequency domain}
In practice, the \gls{rttddft} calculation results in samples of $\delta\rho(t)$ on a finite grid of $N$ times $t_j = j \Delta t$.
We then approximate $\delta \rho_{nn'}(\omega)$ by the discrete Fourier transform
\begin{align}
    \delta \rho_{nn'}(\omega) &\approx \Delta t \textstyle\sum_{j=0}^{N-1} \delta \rho_{ia}(t_j) e^{i\omega t_j}.
\end{align}
A common choice is to perform the \gls{rttddft} calculation using a so-called $\delta$-kick, where $v(t) = K \delta(t)$ and the Fourier transform $v(\omega) = K$ is exactly constant.
For all other kinds of perturbations, we approximate
\begin{align}
    v(\omega) &\approx \Delta t \textstyle\sum_{j=0}^{N-1} v(t_j) e^{i\omega t_j}.
\end{align}
The finite simulation length results in a convolution of the true Fourier transform $\delta \rho_{nn'}(\omega)$ by a sinc-shape, leading to a noisy spectrum.
A common remedy is to artificially dampen the response function, forcing it to zero before the end of the simulation.
In \rhodent{}, Gaussian broadening is implemented, where the delta-kick response is multiplied by a Gaussian envelope of width $\sigma$
\begin{align}
    \delta \rho_{nn'}(\omega; \sigma) &\approx \Delta t \textstyle\sum_{j=0}^{N-1} \delta \rho_{ia}(t_j) e^{-\sigma^2 t_j^2 / 2} e^{i\omega t_j}.
\end{align}
This is equivalent to convoluting the noisy $\delta\rho_{ia}(\omega)$ with a Gaussian $e^{-\omega^2 / 2\sigma^2}$ in the frequency domain.

\subsubsection{In the time domain through the convolution trick}
\label{sec:convolution-trick}
The electron-hole part of the induced density matrix $\rho_{ia}$, where $f_i > f_a$ is linear in the perturbation  \cite{RosErhKui20}, meaning that for sufficiently weak perturbations $\hat{v}$ there is a linear response regime where
\begin{align}
    \delta\rho_{ia}(t) = \int_0^t \hat{\chi}_{ia}(t - t') \hat{v}(t') \mathrm{d}t',
\end{align}
or equivalently, in the frequency domain,
\begin{align}
    \delta\rho_{ia}(\omega) = \hat{\chi}_{ia}(\omega) \hat{v}(\omega),
\end{align}
where $\hat{\chi}_{ia}$ is a response function describing the response of matrix element $\delta\rho_{ia}$ to $\hat{v}$.
By convention, we use indices $i$ and $a$ to denote occupied and unoccupied ground state orbitals (equivalent to holes and electrons), respectively.
Using \rhodent{}, we can exploit this linearity to compute the response $\delta\rho_{ia}'$ to perturbation $\hat{v}'$ knowing the response to $\delta\rho_{ia}$ to perturbation $\hat{v}$, without performing another \gls{rttddft} calculation.
\begin{align}
    \delta\rho_{ia}'(\omega)
    &= \hat{\chi}(\omega)\hat{v}'(\omega) \\
    &= \delta\rho_{ia}(\omega)\frac{v'(\omega)}{v(\omega)}.
    \label{eq:resp:pulseconv}
\end{align}
We are restricted to perturbations of the same spatial shape, where the onset of $v'(t)$ is not earlier in time than $v(t)$, and where the spectrum of $v(\omega)$ covers $v'(\omega)$ entirely.
The response in the time domain can be obtained by carrying out an inverse Fourier transformation of \autoref{eq:resp:pulseconv}.

In practice, quantities are available on a finite grid of $N$ times $t_j = j \Delta t$.
We then take a discrete Fourier transform of the induced density matrix and both perturbations on the grid of $N$ time instances 
\begin{align}
    \delta\hat{\rho}_{k,ia} &= \textstyle\sum_{j=0}^{N-1} \delta \rho_{ia}(t_j) e^{i\omega_k t_j} \\
    \hat{v}_{k} &= \textstyle\sum_{j=0}^{N-1} \delta v(t_j) e^{i\omega_k t_j} \\
    \hat{v}'_{k} &= \textstyle\sum_{j=0}^{N-1} \delta v'(t_j) e^{i\omega_k t_j}.
\end{align}
The chosen frequencies $\omega_k = 2\pi k / (\Delta t N')$ correspond to a grid of $N'$ times ($N'$ is at least $2N$), which is equivalent to padding the data with zeros after the end of the simulation.
This is necessary to prevent circular correlation (see \autoref{sec:circular-conv}).
We obtain the induced density matrix to the new perturbation as the inverse discrete Fourier transform of \autoref{eq:resp:pulseconv}
\begin{align}
    \delta\rho'_{ia}(t_j) = \frac{1}{N'}\sum_{k=0}^{N-1} \frac{\delta\hat{\rho}_{k,ia}\cdot\hat{v}'_k}{\hat{v}_k}e^{-i\omega_k t_j}.
\end{align}

When constructing the density matrix from the time-dependent wave functions file, it is important to write the wave functions at every time step, if a $\delta$-kick was used.
Otherwise, if the wave functions are written at a sparse interval $\Delta t$ there will be aliasing in the Fourier spectrum above the Nyquist frequency $1/2\Delta t$.

\begin{figure}
    \centering
    \includegraphics{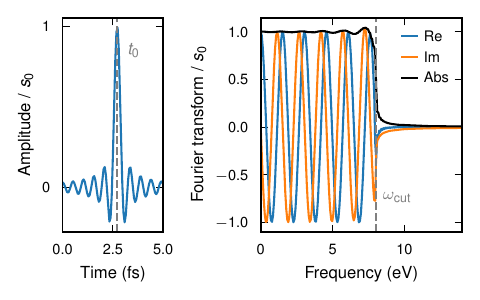}
    \caption{
        Sinc-shaped pulse with cutoff $\hbar\omega_\text{cut} = \qty{8}{\electronvolt}$ and offset $t_0 = (5 + 1/4) / (2\pi\omega_\text{cut})$ in the time and frequency domain.
        Choosing a multiple of an integer plus one quarter ensures that the pulse is finite at the beginning of the simulation.
     }
    \label{fig:sinc-pulse}
\end{figure}

This behavior can be alleviated by using different temporal shapes.
For example, a sinc-shaped pulse (\autoref{fig:sinc-pulse})
\begin{align}
    v(t) = s_0 \frac{\sin(\omega_{cut} (t - t_0))}{\omega_{cut} (t - t_0)}, t > 0
    \label{eq:resp:sinc-pulse}
\end{align}
has the Fourier transform
\begin{align}
    v(\omega) = s_0 \frac{e^{i\omega t_0}}{\omega_\text{cut}}\text{rect}\left(\frac{\omega}{2\pi\omega_\text{cut}}\right),
\end{align}
where the rectangular function $\text{rect}(x)$ is equal to 1 for $0 < x < 1$ and 0 otherwise.
In principle, this pulse does not induce any response above the cutoff frequency and allows saving the time-dependent wave functions file using a sparse interval.

However, in finite simulations, $v(\omega)$ is effectively convoluted with a sinc function leading to a smooth cutoff as $v(t)$ does not fit in the simulation window in its entirety.
Choosing $t_0$, one should strike a balance between fitting as much as possible of $v(t)$ in the simulation window (leading to a sharper cut-off) and perturbing the system with a sufficiently strong field early on in the perturbation (leading to better numerical stability).
In particular, it is a judicious choice to let the offset be an integer and a quarter offset of the time between oscillations $t_0 = (n + 1/4) / (2\pi\omega_\text{cut})$ so that the pulse starts at a local maximum.

\subsubsection{Constructing the KS density matrix}
Currently, \rhodent{} provides an interface to the \gls{lcao} implementation of \gls{rttddft} in \textsc{gpaw} \cite{MorLarKui24, KuiSakRos15, LarVanMor09}.
In this implementation, the \gls{ks} wave functions are represented as the linear combination
\begin{align}
    \psi_n(\vec{r}, t) = \sum_{\mu} C_{n\mu}(t) \Phi_\mu(\vec{r}),
\end{align}
where $\Phi_\mu(\vec{r})$ are atom-centered orbitals (typically a few orbitals for each atom) and only the coefficients $C_{n\mu}$ are time-dependent.

\subsubsection{Time-dependent wave functions file}
In \textsc{gpaw}, the time-dependent coefficients $C_{n\mu}$ can be written to file periodically during time propagation.
From this file, we can construct the \gls{ks} density matrix in the \gls{lcao} basis
\begin{equation}
    \rho_{\mu\nu}(t) =
    \sum_k C_{k\nu}^*(t)f_k C_{k\nu}(t).
    \label{eq:resp:td_wfs_lcao}
\end{equation}
Combining \autoref{eq:resp:rho_nn} and \autoref{eq:resp:td_wfs_lcao}, we find the basis transformation matrix $P_{n\mu} = \sum_{\nu} C_{n\mu}(0) S_{\nu\mu}$ with the overlap integral $S_{\mu\nu} = \int \Phi^*_{\nu}(\vec{r}) \Phi_{\mu}(\vec{r}) \mathrm{d}{\vec{r}}$.
Then the induced density matrix in the basis of ground state \gls{ks} orbitals is
\begin{equation}
    \delta\rho_{nn'}(t) =
    \sum_{\mu\nu} P_{n'\nu}^*\left[\rho_{\mu\nu}(t) - \rho_{\mu\nu}(0)\right] P_{n\mu}.
    \label{eq:resp:td_wfs}
\end{equation}

\subsubsection{Fourier transform of induced KS density matrix. }
\textsc{gpaw} also supports building the Fourier transform $\delta \rho_{\mu\nu}(\omega) = \int_{0}^\infty[\rho_{\mu\nu}(t) - \rho_{\mu\nu}(0)]e^{i\omega t}\mathrm{d}t$ on the fly during time propagation on a grid of predefined frequencies \cite{RosKuiPus17}.
Having saved this quantity to file, \rhodent{} can read it and transform the density matrix to the basis of ground state \gls{ks} orbitals as
\begin{align}
    \rho_{nn'}(\omega) =
    \sum_{\mu\nu} P_{n'\nu}^*\rho_{\mu\nu}(\omega) P_{n\mu}.
\end{align}

\subsubsection{Reading and writing numpy binary files}
The construction of the density matrix from different sources, and transformations between time and frequency domain, are time and memory intensive.
In particular, memory usage can be limiting in parallel execution as, for example, the transformation matrix $P_{n\mu}$ is duplicated on every process.
Therefore, \rhodent{} allows writing the density matrix in the \gls{ks} basis to a NumPy binary file on disk after every transformation.
The files can be read in order to continue with another transformation (for example, the pulse convolution outlined in the next section), or simply to compute observables.

\subsection{Calculating observables}
\label{sec:obs}
In \rhodent{} \texttt{Calculator}s are used to evaluate observables, which can be computed using the full response $\delta\rho_{ia}$, including:
\begin{itemize}
    \item The density in the time domain, or the Fourier transform thereof.
    \item The dipole moment in the time domain, or its Fourier transform (the polarizability in the frequency domain).
    \item Stored energy in the system in the time domain.
    \item Electron and hole occupations, i.e., \gls{hc} distributions, in the time domain.
\end{itemize}
Additionally, a density of states calculator, which only relies on ground state information, and a spectrum calculator, which only needs a time-dependent dipole moment file, are implemented.
For analysis purposes, the observables can be decomposed either by the energies or spatial confinement of occupied and unoccupied states $i$ and $a$.
For the latter, we use the Voronoi cell of an atom, which is the set of all points in space closer to that atom than to any other, to define the Voronoi weights
\begin{align}
    w^\text{proj}_{nn'}
    = \int_\text{Voronoi cell}\psi_{n'}^*(\boldsymbol{r}, 0) \psi_{n}(\boldsymbol{r}, 0) \mathrm{d}\boldsymbol{r}
    \label{eq:obs:voronoi}.
\end{align}
In the following, we use Hartree atomic units.

\subsubsection*{Density}
The induced charge density is obtained from $\delta\rho_{ia}$ as
\begin{align}
    \delta n(\boldsymbol{r}) = - \sum_{ia'} \psi^*_a(\boldsymbol{r}, 0) \psi_i(\boldsymbol{r}, 0) \delta\rho_{ia}.
\end{align}
Because the electron-electron and hole-hole parts of the \gls{ks} density matrix ($\delta\rho_{nn'}$ where $f_n = f_{n'}$) are quadratic in perturbation, the electron-hole part dominates.
Using that $\rho_{ia} = \rho_{ai}^*$ and that the \gls{ks} orbitals are real, we can then write
\begin{align}
    \delta n(\boldsymbol{r}) = -2 \sum_{ia}^{f_i > f_a}
    \psi_a(\boldsymbol{r}, 0) \psi_i(\boldsymbol{r}, 0) \mathrm{Re}\:\delta\rho_{ia}.
\end{align}

An energy filter can be supplied to \rhodent{} to include only a subset of transitions in the sum.
For example, one could want to include only transitions to electrons above $\varepsilon_\text{low}$ in energy
\begin{align}
    \delta n_\text{filter}(\boldsymbol{r}) = -2 \sum_{ia}^{\substack{f_i > f_a \\ \varepsilon_a > \varepsilon_\text{low}}}
    \psi_a(\boldsymbol{r}, 0) \psi_i(\boldsymbol{r}, 0) \mathrm{Re}\:\delta\rho_{ia}.
\end{align}

\subsubsection*{Dipole moment}

The induced dipole moment (dipole moment minus the static part) is
\begin{align}
    \label{eq:obs:dipole_total}
    \delta\boldsymbol{\mu} = -2 \sum_{ia}^{f_i > f_a} \boldsymbol{\mu}_{ia} \mathrm{Re}\:\delta\rho_{ia},
\end{align}
where
\begin{align}
    \boldsymbol{\mu}_{ia} = \int \psi^*_a(\boldsymbol{r}, 0) \boldsymbol{r}\psi_i(\boldsymbol{r}, 0) \mathrm{d}\boldsymbol{r}.
\end{align}

We can use \rhodent{} to calculate the induced dipole \autoref{eq:obs:dipole_total}, the induced dipole projected on Voronoi weights of occupied and unoccupied states for a projection on one or several atoms
\begin{align}
    \label{eq:obs:dm_voronoi}
    \delta\boldsymbol{\mu}^{\text{occ}\rightarrow\text{unocc}, ia} = -2 \sum_{ia}^{f_i > f_a} \boldsymbol{\mu}_{ia} \mathrm{Re}\:\delta\rho_{ia} w^\text{proj,occ}_{ii} w^\text{proj,unocc}_{aa},
\end{align}
or the dipole or projected dipole as a \gls{tcm} \cite{MalLehEnk13}
\begin{align} 
    \label{eq:obs:dipole_tcm}
    -2 \sum_{ia}^{f_i > f_a} 
    & \boldsymbol{\mu}_{ia} \mathrm{Re}\:\delta\rho_{ia} G(\varepsilon_\text{occ} - \varepsilon_i) G(\varepsilon_\text{unocc} - \varepsilon_a) \\
    \begin{split}
    \label{eq:obs:dipole_proj_tcm}
    -2 \sum_{ia}^{f_i > f_a}
    & \boldsymbol{\mu}_{ia} \mathrm{Re}\:\delta\rho_{ia} w^\text{proj,occ}_{ii} w^\text{proj,unocc} \\
    & G(\varepsilon_\text{occ} - \varepsilon_i) G(\varepsilon_\text{unocc} - \varepsilon_a).
    \end{split}
\end{align}
The \gls{tcm} enables illustrative analyses of matrices in the basis of ground state \gls{ks} orbitals, by broadening them onto energy axes using Gaussians
\begin{align}
   G(\varepsilon - \varepsilon_n)
    = \frac{1}{\sqrt{2 \pi \sigma^2}}
    \exp\left(-\frac{
        \left(\varepsilon_i - \varepsilon\right)^2
    }{
        2 \sigma^2
    }\right).
\end{align}

For a perturbation corresponding to a spatially constant electric field polarized in the $x$ direction
\begin{align}
    \hat{v}(t) = v(t) x,
\end{align}
the normalized Fourier transform of the induced dipole moment gives one column of the polarizability
\begin{align}
    \vec{\alpha}_x(\omega) = \frac{\int_{0}^\infty\delta\vec{\mu}(t)e^{i\omega t}\mathrm{d}t}{v(\omega)}.
\end{align}
In order to construct the full polarizability tensor, three \gls{rttddft} calculations with orthogonal polarization directions along $x$, $y$, and $z$ are needed
\begin{align}
    \boldsymbol{\alpha}(\omega) = \begin{bmatrix}
        \vec\alpha_x(\omega) & \vec\alpha_y(\omega) & \vec\alpha_z(\omega)
    \end{bmatrix}.
\end{align}

The optical absorption spectrum, expressed as the oscillator strength function, is related to the polarizability as
\begin{align}
    \label{eq:obs:photoabs}
    S_x(\omega) = \frac{2\omega}{\pi} \mathrm{Im}\:\alpha_{xx}(\omega).
\end{align}
This quantity obeys the sum rule that its integral $\int_0^\infty S(\omega)\mathrm{d}\omega$ equals the number of electrons in the system.
The same kind of decomposition is available in the frequency domain as in the time domain.
For example, we can construct the \gls{tcm} \cite{MalLehEnk13} for the photoabsorption cross section
\begin{align}
    \label{eq:obs:tcm_photoabs}
    -\frac{4\omega}{\pi} \sum_{ia}^{f_i > f_a} \boldsymbol{\mu}_{ia}\frac{\text{Im}\:\left[\mathrm{Re}\:\delta\rho_{ia}\right](\omega)}{v(\omega)}G(\varepsilon_\text{occ} - \varepsilon_i) G(\varepsilon_\text{unocc} - \varepsilon_a)
\end{align}

\subsubsection*{Stored energy}
In \gls{ks} \gls{dft}, the total energy $E_\text{tot}$ can be partitioned into the kinetic energy $T$, the potential energy due to the Hartree and \gls{xc} potential $E_\text{Hxc}$, and the external energy.
Using perturbation expansions up to second order \cite{RosErhKui20} we can write the decomposition of the energy stored in the system
\begin{align}
    E_\text{tot}(t) = E_\text{tot}(0) + \delta T(t) + \delta E_\text{Hxc}(t) + E_\text{field}(t),
\end{align}
where the first term is the total energy of the ground state and the last term is the external energy due to the electric field of the perturbation
\begin{equation}
    \label{eq:obs:energy_field}
    E_\text{field}(t) = - \delta\vec{\mu}(t)\cdot \hat{\boldsymbol{e}} v(t).
\end{equation}
Here, we assume that the electric field of the perturbation is spatially constant and polarized in $\hat{\boldsymbol{e}}$ direction.
The change in kinetic and \gls{hxc} energies can be decomposed into contributions from all electron-hole transitions
\begin{align}
    \label{eq:obs:energy_total}
    \delta T(t) + \delta E_\text{Hxc}(t) &= \sum_{ia}^{f_i > f_a} E_{ia}  \\
    \label{eq:obs:energy_hxc}
    \delta E_\text{Hxc}(t) &= \sum_{ia}^{f_i > f_a} E_{ia}^\text{Hxc},
\end{align}
where
\begin{align}
    E_{ia} &= \frac{1}{2} \left[
    p_{ia}\dot{q}_{ia} - q_{ia} \dot{p}_{ia} - v_{ia} q_{ia} \right] \\
    E_{ia}^\text{Hxc} &= -\frac{1}{2} \left[
    \omega_{ia} q_{ia}^2 - q_{ia} \dot{p}_{ia} - v_{ia} q_{ia} \right]
\end{align}
are calculated from the following quantities and their time derivatives
\begin{align}
    p_{ia} &= \frac{2\:\mathrm{Im}\:\delta\rho_{ia}}{\sqrt{2 f_{ia}}} \\
    q_{ia} &= \frac{2\:\mathrm{Re}\:\delta\rho_{ia}}{\sqrt{2 f_{ia}}} \\
    f_{ia} &= f_a - f_i \\
    v_{ia} &= \sqrt{2 f_{ia}}
    \boldsymbol{\mu}_{ia} \cdot\hat{\boldsymbol{e}}
    v(t).
\end{align}

We can also calculate the rate of energy change using higher derivatives
\begin{align}
    \dot{E}_{ia} = \frac{1}{2} \left[
    p_{ia}\ddot{q}_{ia} - q_{ia} \ddot{p}_{ia}
    - v_{ia} \dot{q}_{ia} - \dot{v}_{ia} q_{ia} \right].
\end{align}

For energies, we can compute Voronoi projections analogously to \autoref{eq:obs:dm_voronoi}, and perform energy filtering based on the energy of pairs $ia$.
For example, we can compute the total and \gls{hxc} energy above a threshold $\varepsilon_\text{low}$
\begin{align}
    \delta E_\text{tot}(t) &= \sum_{ia}^{\substack{f_i > f_a \\ \varepsilon_a - \varepsilon_i > \varepsilon_\text{low}}} E_{ia} \\
    \delta E_\text{Hxc}(t) &= \sum_{ia}^{\substack{f_i > f_a \\ \varepsilon_a - \varepsilon_i > \varepsilon_\text{low}}} E_{ia}^\text{Hxc}.
\end{align}
Similarly, we can compute \glspl{tcm} analogously to \autoref{eq:obs:dipole_tcm} and \autoref{eq:obs:dipole_proj_tcm}.

\subsubsection*{Hot carriers}
The equal-occupation-numbers part ($f_{n} = f_{n'}$) of the induced density matrix  is quadratic in the perturbation \cite{RosErhKui20}, and can be obtained from the electron-hole part
\begin{equation}
    \delta\rho_{nn'}(t) = \delta\rho^\text{electrons}_{nn'}(t) - \delta\rho^\text{holes}_{nn'}(t),
    \label{eq:obs:second_order}
\end{equation}
where 
\begin{align}
    \delta\rho^\text{holes}_{ii'} &= 
        \frac{1}{2} \sum_a^{f_i > f_a}
        p_{i'a} p_{ia} + q_{ia} q_{i'a} \\
    \delta\rho^\text{electrons}_{aa'} &=
        \frac{1}{2} \sum_i^{f_i > f_a}
        p_{ia} p_{ia'} + q_{ia} q_{ia'}.
\end{align}
The purely electron-electron ($\delta\rho_{aa'}$, $f_a = f_{a'} = 2$) and hole-hole parts ($\delta\rho_{ii'}$, $f_i = f_{i'} = 0$) of the \gls{ks} density matrix are equal to $\delta\rho^\text{electrons}_{aa'}$ and $\delta\rho^\text{holes}_{aa'}$, respectively, but for the partially occupied states $0 < f_n = f_{n'} < 2$ both terms in \autoref{eq:obs:second_order} are non-zero.

Hot-carrier distributions are calculated by convolution of $\delta\rho_{ii'}$ (holes)
and $\delta\rho_{aa'}$ (electrons) with Gaussian broadening functions on the energy
grid.
\begin{align}
    \label{eq:obs:holes}
    P^\text{holes}(\varepsilon) &=
    \sum_i \delta\rho_{ii} G(\varepsilon - \varepsilon_i) \\
    \label{eq:obs:electrons}
    P^\text{electrons}(\varepsilon) &=
    \sum_a \delta\rho_{aa} G(\varepsilon - \varepsilon_a).
\end{align}
We can also project hot-carrier distributions according to the Voronoi weights
\begin{align}
    \label{eq:obs:holes_proj}
    P^\text{proj,holes}(\varepsilon) &=
    \sum_{ii'} \delta\rho_{ii'} w^\text{proj}_{ii'} G(\varepsilon - \varepsilon_i) \\
    \label{eq:obs:electrons_proj}
    P^\text{proj,electrons}(\varepsilon) &=
    \sum_{aa'} \delta\rho_{aa'} w^\text{proj}_{aa'} G(\varepsilon - \varepsilon_a).
\end{align}

\subsubsection*{Optical spectrum and densities of state}

Optical response calculations are a routine task in \gls{rttddft} since the calculations by Yabana and Bertsch \cite{YabBer96}.
For spectroscopic applications, the dipolar response of the matter system due to a dipolar external field is of interest, as it is the only non-vanishing contribution in the far field.

\begin{align}
    \vec{d}(\omega) = \vec{\alpha}(\omega) \vec{E}_\text{ext}(\omega),
\end{align}
or, equivalently, in the time domain
\begin{align}
    \vec{d}(t) = \int_0^t \vec{\alpha}(t - t') \vec{E}_\text{ext}(t') \mathrm{d}t'.
\end{align}

Finally, we can compute the total \gls{dos}
\begin{align}
    \label{eq:obs:dos}
    \sum_k G(\varepsilon - \varepsilon_k)
\end{align}
and the \gls{pdos} as
\begin{align}
    \label{eq:obs:pdos}
    \sum_k G(\varepsilon - \varepsilon_k) w^\text{proj}_{kk},
\end{align}
where $w^\text{proj}_{kk}$ are the Voronoi weights given by \autoref{eq:obs:voronoi}.

\section{Examples}

In this section, we study three plasmonic systems to demonstrate the capabilities of \rhodent{}.
The calculation workflow is as follows:
For each system, we first compute the electronic ground state within \gls{dft} using \gls{lcao} basis sets in \textsc{gpaw}.
Then we perform an \gls{rttddft} calculation using \textsc{gpaw}, propagating the ground state in time under an external perturbation.
We simulate the dynamics for \qty{30}{\femto\second}, which is enough to resolve the formation and subsequent dephasing of the \gls{lspr}.
A propagation time step of \qty{10}{\atto\second} yields converged results.
We apply a sufficiently weak perturbation to only probe the linear-response domain.
By choosing a sinc-shaped (\autoref{eq:resp:sinc-pulse}) perturbation in the time domain, we avoid exciting high frequency components of the response, which allows us to write a wave function trajectory to file at a relatively sparse interval, without risking aliasing.
This file is read by \rhodent{} to calculate the response.
Further information concerning the \gls{dft} and \gls{rttddft} calculations can be found in \autoref{sec:comp-details}.

\subsection{Aluminum nanoparticle}

As an initial example, we consider a plasmonic 201-atom Al \gls{np}, with a diameter of \qty{1.6}{\nano\meter}.
An in-depth analysis of this system can be found in Ref.~\cite{RosSheErh19}.
We perform one \gls{rttddft} calculation, using a sinc-pulse with a cutoff of \qty{16}{\electronvolt}, and write the wave function trajectory to disk at an interval of \qty{100}{\atto\second} (Nyquist frequency of \qty{20.7}{\electronvolt}).
Due to the symmetry of the \gls{np}, its response is isotropic, and one polarization direction is enough to probe the entire response.

\subsubsection*{Analysis in the frequency domain}

\begin{figure}
    \centering
    \includegraphics{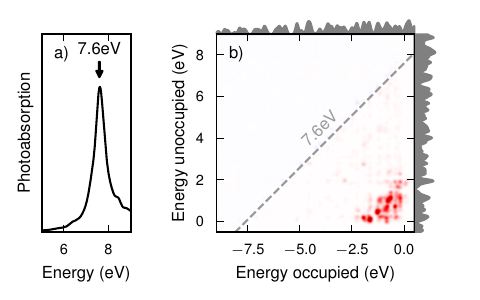}
    \caption{
    (a) Absorption spectrum of the Al \gls{np}.
    (b) \Glsdesc{tcm} (\glsfirst{tcm}) for the absorption of the Al \gls{np} at the frequency $\hbar\omega = \qty{7.6}{\electronvolt}$ (indicated in panel (a)).
    Energies are given with respect to the Fermi level.
    The diagonal line is drawn over resonant electron-hole transitions, i.e., those with an energy difference equal to $\hbar\omega$.
    There are no significant contributions to photoabsorption from resonant transitions.
    Instead, the main contributions are from transitions near the Fermi level.
    }
    \label{fig:Al-tcm}
\end{figure}

Using the dipole moment file from the \gls{rttddft} calculation, and \autoref{eq:obs:photoabs}, we construct a photoabsorption spectrum (\autoref{fig:Al-tcm}a) for the system.
The main feature in the spectrum is a peak at \qty{7.6}{\electronvolt}.
We attribute the peak to the \gls{lspr}, which is a collective electronic excitation \cite{Boh83, HutFen04}.
We can gain insight about the microscopic origin of this feature by constructing the \gls{tcm} to the absorption, according to \autoref{eq:obs:tcm_photoabs}.
This calculation requires the full response of the system, i.e., the wave function trajectory.
Visualizing the contributions as a \gls{tcm} (\autoref{fig:Al-tcm}b), we see that many individual electron-hole pairs near the Fermi level collectively contribute to the absorption.
We also compute the \gls{dos}, using \autoref{eq:obs:dos}, and plot it together with the \gls{tcm}.
Despite the energy difference between most involved electron and hole states being less than 3 to \qty{4}{\electronvolt} or so, they are excited at \qty{7.6}{\electronvolt}, thanks to their collective coupling through Coulomb interactions.

\subsubsection*{Analysis in the time domain}

\begin{figure}
    \centering
    \includegraphics{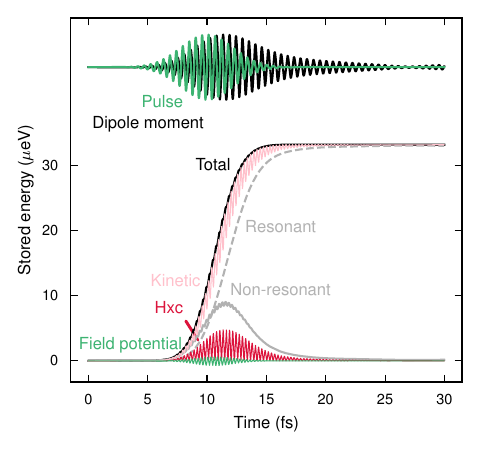}
    \caption{Time-evolution of energy stored in the Al \gls{np} excited by a laser pulse resonant to the \gls{lspr}.
    The total energy is decomposed into \gls{hxc} contributions and kinetic contributions.
    An alternative decomposition is into energy of resonant and non-resonant transitions.
    The potential energy due to the external electric field (not included in the total), the pulse, and the induced dipole moment are also plotted in the figure.
    }
    \label{fig:Al-energy}
\end{figure}

\begin{figure*}
    \centering
    \includegraphics{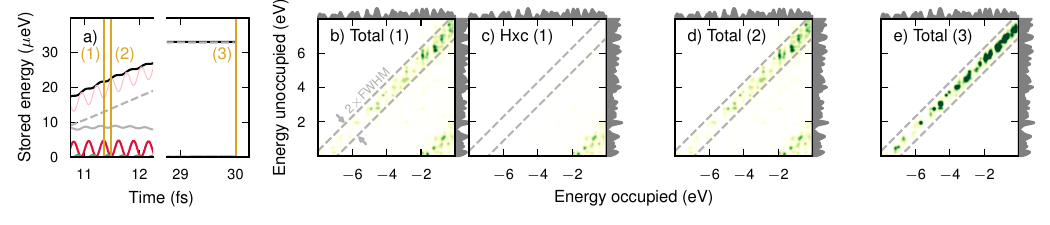}
    \caption{(a) Time-evolution of energy stored in the Al \gls{np} excited by a laser pulse resonant to the \gls{lspr}, with time instances (1), (2) and (3) marked.
    (b) \gls{tcm} of total and of (c) \gls{hxc} contributions at time instance (1). 
    (d) \gls{tcm} of total stored energy at time instance (2).
    (e) \gls{tcm} of total stored energy at time instance (3).
    The transitions counted as resonant are indicated by dashed lines in (b-e). 
    The same normalization is used in all axes for the colormap.
    }
    \label{fig:Al-energy-tcm}
\end{figure*}

Now we consider the response of the Al \gls{np} to a Gaussian laser pulse.
We consider a spatially constant electric field
\begin{equation}
    \vec{E}_\text{pulse}(t) = \vec{E}_0 \cos(\omega_0 (t - t_0)) \exp(-(t-t_0)^2/\tau^2),
\end{equation}
choosing the strength $|\vec{E}_0|=\qty{51}{\mu\volt/\angstrom}$, an envelope centered at $t_0=\qty{10}{\femto\second}$, and $\tau = \qty{2.19}{\femto\second}$, giving a \gls{fwhm} of about \qty{5.2}{\femto\second} in the time domain and \qty{0.7}{\electronvolt} in the frequency domain. 
Thanks to the pulse convolution functionality in \rhodent{} (see \autoref{sec:convolution-trick}), we do not need to perform any new \gls{rttddft} calculations, which is the most expensive part of the simulation workflow.
By appropriately setting up the \texttt{Response} object, the pulse convolution is performed before calculating observables.

We excite the system at the \gls{lspr} ($\hbar\omega_0 = \qty{7.6}{\electronvolt})$, and evaluate the time-dependence of the energy stored in the system (\autoref{fig:Al-energy}).
As the pulse is turned on, the system responds with a dipole moment $\mu$ that lags roughly $\pi/2$ in phase after the electric field, corresponding to resonant response (absorption), and peaks in amplitude a few oscillations after the maximum of the pulse.
Therefore, the energy contribution associated with the pulse (\autoref{eq:obs:energy_field}) oscillates around zero at twice the pulse frequency.

The total energy increases during the entire duration of the pulse until it reaches a steady value.
Having access to the full response, we can partition the total energy into kinetic and \gls{hxc} contributions, $\delta T(t)$ and $\delta E_\text{Hxc}(t)$, see \autoref{eq:obs:energy_total} and \autoref{eq:obs:energy_hxc}.
We can see how energy is periodically redistributed between the former and the latter, at a frequency matching the pulse.
Increasingly more energy is stored in kinetic contributions, while the amount of \gls{hxc} energy periodically reaches zero.
The \gls{hxc} contributions persist for a few \qty{}{\femto\second} after the disappearance of the pulse, so we interpret this as the lifetime of the \gls{lspr}.
Alternatively, we can partition the total energy into resonant and non-resonant contributions, by the distinction that the eigenvalue difference of resonant electron-hole pairs should be $|\hbar\omega_{ia} - \hbar\omega_0| < \qty{1.4}{\electronvolt}$.
This tells us that the number of non-resonant transitions rises and falls smoothly, with only small oscillations, during the lifetime of the \gls{lspr}.

For even deeper insight into the data, we can construct \glspl{tcm} for the energy contributions at selected times (\autoref{fig:Al-energy-tcm}).
At the time instance marked (1) in \autoref{fig:Al-energy-tcm}a, when the \gls{hxc} contributions are at a maximum, we see both resonant and non-resonant transitions in the total energy (\autoref{fig:Al-energy-tcm}b), and only non-resonant transitions in the \gls{hxc} contribution (\autoref{fig:Al-energy-tcm}c).
After one pulse cycle, at time instance (2), there are no \gls{hxc} contributions, but the same non-resonant transitions persist in the total energy (\autoref{fig:Al-energy-tcm}d).
We should note here that these are the same transitions that are visible in the absorption \gls{tcm} (\autoref{fig:Al-tcm}b).
It is now clear that these transitions make up the \gls{lspr}, and that they oscillate between kinetic and potential energy during its lifetime.
After the decay of the pulse, the plasmon dephases into resonant transitions, known as \glspl{hc}.
At time instance (3) only such transitions remain (\autoref{fig:Al-energy-tcm}e).

\subsection{Aluminum nanoparticle with benzene molecules}

The next example system is another structure from Ref.~\cite{RosSheErh19}, made up of an Al \gls{np} identical to the previous structure, and two benzene molecules.
The molecules are placed on opposite sides at a distance of \qty{3}{\angstrom} from the faces of the \gls{np}.
The molecules are oriented such that the line through the molecules and the \gls{np} is parallel with the transition dipoles of the lowest bright excitations of the molecules.
The system is no longer isotropic, so we constrain our analysis to excitation polarized along this axis.
We perform a new \gls{rttddft} calculation, with the same parameters as for the bare Al \gls{np}.

\subsubsection*{Analysis in the frequency domain}

\begin{figure}
    \centering
    \includegraphics{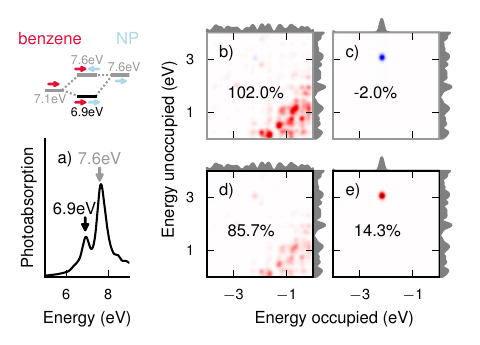}
    \caption{(a) Absorption spectrum of the Al \gls{np} and benzene system.
    (b-e) \Glspl{tcm} for the absorption, for subset of transitions:
    (b) to the \gls{np}, at the \gls{up} frequency \qty{7.6}{\electronvolt},
    (c) to the molecule, at the \gls{up} frequency \qty{7.6}{\electronvolt},
    (d) to the \gls{np}, at the \gls{lp} frequency \qty{6.9}{\electronvolt},
    (e) to the molecule, at the \gls{lp} frequency \qty{6.9}{\electronvolt}.
    For each of (b--e), the sum of the subset of transitions is written out as a fraction of the sum of all transitions at the same frequency.
    }
    \label{fig:Al-mol-tcm}
\end{figure}

We calculate the photoabsorption spectrum (\autoref{fig:Al-mol-tcm}a), and find two peaks.
One peak at \qty{7.6}{\electronvolt} is very similar to the \gls{lspr} peak of the bare Al \gls{np}, but the other at \qty{6.9}{\electronvolt} is both much larger in prominence and redshifted by about \qty{0.2}{\electronvolt} compared to the peak of the bare molecules.
This is a characteristic of strong coupling, where the underlying excitations (\gls{lspr} of the \gls{np} and the molecular excitation) exchange energy with each other at a faster or comparable rate to the lifetime of the excitations \cite{TorBar14}.
The two resonances are known as the \gls{lp} and \gls{up}.

We decompose the absorption \glspl{tcm} into \gls{np}-like and molecular transitions, using the Voronoi decomposition \autoref{eq:obs:dm_voronoi}.
In particular, we attribute transitions from occupied \gls{np}-states into any unoccupied state as the former, and transitions from occupied molecular states to any unoccupied state as the latter, because unoccupied states tend to be more of mixed character compared to occupied states.
In \autoref{fig:Al-mol-tcm}b--e, we plot the projected \glspl{tcm} together with the \gls{dos} (\autoref{eq:obs:dos}) on the unoccupied energy axes, and the \gls{pdos} (\autoref{eq:obs:pdos}) for the \gls{np}/molecule on the occupied energy axes.
At the \gls{up} frequency $\hbar\omega_0 = \qty{7.6}{\electronvolt}$, we find that the \gls{np} decomposition of the \gls{tcm} (\autoref{fig:Al-mol-tcm}b) consists of the same excitations near the Fermi level that we attributed to the \gls{lspr} in the bare Al \gls{np}.
Meanwhile, the molecular decomposition of the \gls{tcm} (\autoref{fig:Al-mol-tcm}c) consists of a single excitation from the benzene \gls{homo} to the \gls{lumo}.
The molecular transition is contributing destructively to the absorption at the frequency of the \gls{up}, meaning that the molecular dipole is directed opposite to the \gls{np} dipole.
At the \gls{lp} frequency $\hbar\omega_0 = \qty{6.9}{\electronvolt}$, the \gls{tcm} also consists of plasmonic transitions in the \gls{np} decomposition (\autoref{fig:Al-mol-tcm}d) and the single molecular excitation in the molecular decomposition (\autoref{fig:Al-mol-tcm}e).
The difference is that for the \gls{lp}, the \gls{lspr} and molecular excitation couple constructively.
A configuration where the dipoles are parallel intuitively leads to a lower energy.
The \gls{lp} has significant molecular character, as \qty{14.3}{\%} of the absorption comes from the molecular transition, while the \gls{up} is almost entirely \gls{lspr}-like (the destructive contribution of the molecule is only \qty{2.0}{\%}).

\subsubsection*{Analysis in the time domain}

\begin{figure}
    \centering
    \includegraphics{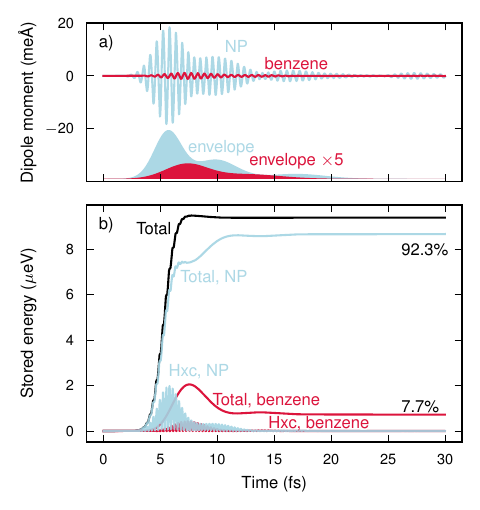}
    \caption{
        For the Al \gls{np} and benzene molecule excited between the polaritons at frequency \qty{7.25}{\electronvolt}, the time-evolution of (a) the induced dipole moment (decomposed into transitions to the \gls{np} and molecule) and (b) stored energy (decomposed into total/\gls{hxc} contributions and transitions to \gls{np}/molecule).
        The fraction of the total energy stored in the \gls{np} and molecule respectively, at the end of the simulation, are written out.
        An envelope of three/two Gaussians $A e^{-(t-t_0)^2/2\tau^2}$ has been fitted to the dipole moment of the \gls{np}/molecule, shown as shaded areas in (a).
    }
    \label{fig:Al-mol-energy}
\end{figure}

Now we consider the time evolution of the dipole moment and energies after excitation by a Gaussian laser pulse.
We excite the system with a pulse covering both peaks, with parameters $\hbar\omega_0 = \qty{7.25}{\electronvolt}$, $t_0=\qty{5}{\femto\second}$, and $\tau = \qty{1.10}{\femto\second}$ (giving a \gls{fwhm} of about \qty{2.6}{\femto\second} in the time domain and \qty{1.4}{\electronvolt} in the frequency domain).
We decompose the energies and dipoles in the same way as for the earlier frequency domain analysis, and find that a strong oscillating dipole first forms in the \gls{np}, then in the molecules, and then in the \gls{np} again (\autoref{fig:Al-mol-energy}).
These are Rabi oscillations, with 2-\qty{3}{\femto\second} in between the maxima.
Oscillations in the \gls{hxc} energy are visible in the \gls{np} and molecules during the same times as oscillations in the dipole.
After \qty{20}{\femto\second} of simulation time, only kinetic energy remains (in the form of \glspl{hc}), with a 92.3/\qty{7.7}{\%} division between the \gls{np} and molecules.

\subsection{Silver nanoparticle with CO molecule}

As a final example, we consider a metal with valence d-electrons.
We set up a 201-atom Ag \gls{np} (nearly the same diameter and shape as the Al \gls{np}),
and place a CO molecule at a distance of \qty{3}{\angstrom} from the (111) on-top site.
Detailed analysis of this system can be found in Ref.~\cite{FojRosKui22}  
We perform yet another \gls{rttddft} calculation, with a polarization direction along the bond axis of the molecule.
As for the Al systems, we use a time step of \qty{10}{\atto\second} and a total simulation length of \qty{30}{\femto\second}.
Because the relevant resonances are lower for this system compared to the Al \gls{np}, we use a sinc-pulse with a cutoff of \qty{8}{\electronvolt}.
This allows us to write the wave function trajectory to disk at an interval of \qty{200}{\atto\second} (Nyquist frequency of \qty{10.3}{\electronvolt}).

\subsubsection*{Analysis in the frequency domain}

\begin{figure}
    \centering
    \includegraphics{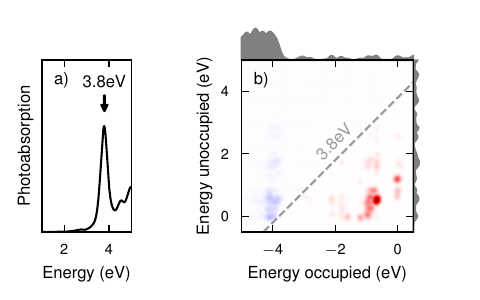}
    \caption{
    (a) Absorption spectrum of the Ag \gls{np} and CO molecule.
    For this system, the molecule does not contribute to the absorption significantly.
    (b) \Glsdesc{tcm} (\glsfirst{tcm}) for the absorption of the same system.}
    \label{fig:Ag-tcm}
\end{figure}

We calculate the photoabsorption spectrum (\autoref{fig:Ag-tcm}a) and identify the \gls{lspr} peak at \qty{3.8}{\electronvolt}.
Unlike the Al--benzene system we studied previously, the first excitation of the CO molecule is far from the \gls{lspr} and its presence has a negligible effect on the absorption spectrum.
We construct a \gls{tcm} for the absorption spectrum (\autoref{fig:Ag-tcm}b) and see many excitations near the Fermi level constructively coupling at the \gls{lspr} frequency.
A new feature, not found in the Al systems, are the destructive contributions involving transitions from the d-band edge (about \qty{4}{\electronvolt} below the Fermi level) to unoccupied levels.
These transitions are excited at the same time as the \gls{lspr}, with their dipoles opposite.
Thus they screen the plasmon, contributing to a much lower resonance energy (\qty{3.8}{\electronvolt}) compared to the Al \gls{np} (\qty{7.6}{\electronvolt}).

\subsubsection*{Analysis in the time domain}

\begin{figure}
    \centering
    \includegraphics{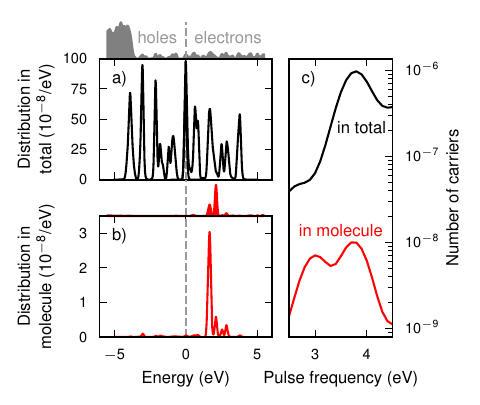}
    \caption{(a) Hot-carrier distribution for the Ag \gls{np} + CO molecule system after excitation at resonance (\qty{3.8}{\electronvolt}).
    The \gls{dos} is drawn above the axes.
    (b) Hot-carrier distribution projected on the molecule after excitation at resonance (\qty{3.8}{\electronvolt}).
    The \gls{pdos} of the molecule is drawn above the axis.
    (c) The total number of excited electrons, and number of excited electrons in the molecule, as a function of excitation energy.
    }
    \label{fig:Ag-hc}
\end{figure}

Once again, we calculate the response to a Gaussian laser pulse.
We choose the parameters $\hbar\omega_0 = \qty{3.8}{\electronvolt}$, $t_0=\qty{10}{\femto\second}$, and $\tau = \qty{2.19}{\femto\second}$ (\gls{fwhm} \qty{7.3}{\femto\second} in the time domain and \qty{0.7}{\electronvolt} in the frequency domain) in order to cover the \gls{lspr} peak.
For this system, we focus on the time after the dephasing of the plasmon.
We compute the average of the \gls{hh} and \gls{he} distributions (\autoref{fig:Ag-hc}a) during the last \qty{5}{\femto\second} of the simulation, using \autoref{eq:obs:holes} and \autoref{eq:obs:electrons}.
We see that \glspl{hc} are created in the \gls{np}, with energies spanning from 0 to the pulse frequency (or negative pulse frequency, for holes).
As we saw in the energy \gls{tcm} for the Al-system (\autoref{fig:Al-energy-tcm}e), \glspl{hc} present after the dephasing of the plasmon are found on the diagonal $\varepsilon_a - \varepsilon_i = \hbar\omega_0$, where $\varepsilon_a$ and $\varepsilon_i$ are the energies of the electron and hole respectively.
We also see this in that the \gls{he} distribution looks like the \gls{hh} distribution, shifted upwards in energy by the pulse energy, except for some minor shifts due to the finite width of the pulse.
At the Fermi level, there are both excited electrons and holes, due to the presence of partially occupied states in the ground state calculation.

For the Ag \gls{np} + CO molecule system, the molecular levels are hybridized with the metal states.
The \gls{lumo} is split into several levels, around 1 to \qty{2}{\electronvolt} above the Fermi level (\autoref{fig:Ag-hc}b; top axes), while the \gls{homo} is too far below the Fermi level to allow for excitations at the present pulse energy.
Therefore, we can expect some \glspl{he} to form in the molecule but not any \glspl{hh}.
Indeed, computing the \gls{hc} distribution projected on the molecule (\autoref{eq:obs:holes_proj} and \autoref{eq:obs:electrons_proj}) confirms that electrons are excited in the molecule in all branches of the hybridized \gls{lumo} (\autoref{fig:Ag-hc}b).
Interestingly, there is a delicate interplay between the coupling of each possible \gls{hc} to the \gls{lspr} and the energetic alignment to the pulse that determines in which hybridized level most electrons are excited.
This has been exploited in \cite{FojRosKui22} to increase the amount of charge transfer.

Using \rhodent{}, we can, without much additional effort, compute the \gls{hc} distributions for many different pulses at once.
In \autoref{fig:Ag-hc}c, we plot the total number of excited electrons in the system, as well as the number of excited electrons in the molecule, by pulse frequency.
The total number of electrons is related to the amount of energy absorbed, which can be deduced from the absorption spectrum pulse frequency roughly like the absorption spectrum, as all the energy absorbed turns into \glspl{hc}.
For the number of electrons in the molecule, there are, however, several local maxima.

\section{Conclusions and outlook}

\rhodent{} is a Python package for post-processing the response of \gls{rttddft} calculations.
It can be used to obtain, e.g., induced dipole moments and densities, the stored energy and \gls{hc} distributions, as well as spatial and energetic projections thereof.
Induced dipole moments and densities can also be computed in the frequency domain (the former being related to optical spectra).
Additionally, since \gls{rttddft} probes the linear response function when the external perturbation is sufficiently weak, \rhodent{} can quickly compute the response to any time-dependent perturbation in the linear response regime, from just one \gls{rttddft} calculation.
In the linear response regime, we can also exploit the structure of the density matrix to compute energies and \gls{hc} distributions in the system.

Currently, rhodent supports reading output files from calculations done using \textsc{gpaw}.
However, thanks to the modular structure of the \rhodent{} code, with a clear separation between the construction of response and the computation of observables, support for other \gls{rttddft} softwares can be readily added.
On a similar note, only calculations done in the \gls{lcao} basis are currently supported.
For \gls{lcao} calculations, we can transform the \gls{ks} density matrix to the basis of ground state \gls{ks} orbitals without any loss of information, and still have a quantity of tractable size.
For, e.g., uniform grid or plane-wave based calculations, the number of basis functions is much larger, and the basis of \gls{ks} orbitals would have to be truncated.

\section{Computational details}

\label{sec:comp-details}

The open-source \textsc{gpaw} \cite{MorLarKui24} code package was used for the \gls{dft} and \gls{rttddft} calculations.
The calculations were done in the \gls{paw} \cite{Blo94} formalism using \gls{lcao} basis sets \cite{LarVanMor09}; the \emph{pvalence} \cite{KuiSakRos15} basis set, which is optimized to represent bound unoccupied states, was used for the Ag species, and the double-zeta polarized (dzp) basis set for all other species.
The PBE \cite{PerBurErn96, PerBurErn97} (for the Al and Al + benzene calculations) and \gls{gllbsc} \cite{GriLeeLen95, KuiOjaEnk10} (for the Ag + CO calculations) \gls{xc}-functionals, utilizing the Libxc \cite{LehSteOli18} library, were used in \textsc{gpaw}.
For the Al and Al + benzene calculations, a simulation cell of \qtyproduct{28.8 x 28.8 x 43.2}{\angstrom} was used.
For the Ag + CO calculations, a simulation cell of \qtyproduct{32.0 x 32.0 x 35.2}{\angstrom} was used.
In the simulation cell, wave functions were represented with a grid spacing of \qty{0.2}{\angstrom}, and \gls{xc} and Coulomb potentials with a grid spacing of \qty{0.1}{\angstrom}.
Additional analytic moment corrections \cite{CasRubSto03} centered at the \gls{np} were added to the Coulomb potential.
Fermi-Dirac occupation number smearing with width \qty{0.05}{\electronvolt} was used.
The self-consistent loop was stopped when the integral of the difference between two subsequent densities was less than \qty{1e-12}{}.
Pulay \cite{Pul80}-mixing was used to accelerate the ground state convergence.

\section*{Data availability}

The data generated for the examples in this article is available on Zenodo at \url{https://doi.org/10.5281/zenodo.16746428}.
The \rhodent{} releases are accessible on Zenodo at \url{https://doi.org/10.5281/zenodo.13332634} while the documentation can be accessed at \url{https://rhodent.materialsmodeling.org}.

\section*{Acknowledgments}

We acknowledge funding from the Knut and Alice Wallenberg foundation (Grant No.~2019.0140; J.~F. and P.~E.) and the Swedish Research Council (No.~2020-04935; J.~F. and P.~E.).
The computations were enabled by resources provided by the National Academic Infrastructure for Supercomputing in Sweden (NAISS) at NSC, PDC and C3SE partially funded by the Swedish Research Council through grant agreement no. 2022-06725.

\appendix

\section{The convolution trick in finite simulations}
\label{sec:circular-conv}

In the linear response regime, we have that (dropping indices $ia$ from the notation in the main text)
\begin{align}
    \delta\rho(t)
    &= \int_0^t \chi(t') v(t - t') \mathrm{d}t',
    \label{app:lin_res_time}
\end{align}
where we assume that the perturbation $v(t)$ is zero before time zero, and the response is causal $\chi(t) = 0$ for $t<0$.
We can then work in the frequency domain
\begin{align}
    \delta\rho(\omega) = \chi(\omega)v(\omega),
\end{align}
where the Fourier transform is defined
\begin{align}
    \delta\rho(\omega) = \int_{-\infty}^{\infty} \delta\rho(t)e^{i\omega t} \mathrm{d}t,
\end{align}
and likewise for $\chi(\omega)$ and $v(\omega)$.
As the Fourier transform is defined as an integration with infinite limits, and our simulations are finite, we need to reformulate the above relations in terms of finite integrations.
Here, we derive a frequency domain formula for quantities sampled in the time window $0 < t < T$.

First, we define the zero-padded quantity
\begin{align}
    v^{(0)}(t) &= \begin{cases}
        v(t) &, 0 < t < T \\
        0 &, T < t < T'
    \end{cases},
\end{align}
which we take to be periodic with the period $T'$.
Next, the crucial assumption is that $T'$ is at least twice the length of $T$.
Then, for $0 < t < T$, we can swap the perturbation by the zero-padded perturbation, and extend the limits of the convolution \autoref{app:lin_res_time}
\begin{align}
    \delta\rho(t) = \int_0^{T} \chi(t') v^{(0)}(t - t') \mathrm{d}t'.
    \label{app:lin_res_extended}
\end{align}
This holds because the added integral is zero
\begin{align}
    \int_t^{T} \chi(t')\underbrace{v^{(0)}(t - t')}_{=0}  \mathrm{d}t' = 0.
\end{align}

We can now expand the zero-padded perturbation in a Fourier series of periodicity $T'$
\begin{align}
    v^{(0)}(t) = \frac{1}{T'}\sum_{k=-\infty}^\infty \hat{v}^{(0)}_k e^{-i\omega_k t},
    \label{app:v_fourier_series}
\end{align}
where $\omega_k = 2\pi k / T'$ and the Fourier coefficients
\begin{align}
    \hat{v}^{(0)}_k =
    \int_0^{T'}v^{(0)}(t) e^{i\omega_k t}\mathrm{d}t
    = \int_0^{T}v(t) e^{i\omega_k t}\mathrm{d}t
\end{align}
only require integration over the true perturbation in the window $0 < t < T$.
Inserting the expansion of $v^{(0)}$ into \autoref{app:lin_res_extended}, we get
\begin{align}
    \delta\rho(t) &=\frac{1}{T'}
    \sum_{k=-\infty}^\infty
    \hat{v}^{(0)}_k e^{-i\omega_k t}
    \int_{0}^{T} \chi(t') e^{i\omega_k t'} \mathrm{d}t' \\
    &= \frac{1}{T'}\sum_{k=-\infty}^\infty
    \hat\chi^{(0)}_k \hat{v}^{(0)}_k e^{-i\omega_k t},
    \label{app:rho_fourier_series}
\end{align}
where we have defined the Fourier coefficients of the zero-padded response function
\begin{align}
    \hat{\chi}^{(0)}_k = \int_0^{T}\chi(t) e^{i\omega_k t}\mathrm{d}t.
\end{align}
We identify \autoref{app:rho_fourier_series} as a Fourier series for $\delta\rho(t)$, valid in the simulation time window $0 < t < T$, with the Fourier coefficients
\begin{equation}
    \delta\hat\rho_k = \hat{\chi}^{(0)}_k\hat{v}^{(0)}_k.
\end{equation}
These coefficients are not, in general, equal to the coefficients of the zero-padded induced density matrix
\begin{align}
    \delta\hat{\rho}^{(0)}_k = \int_0^{T}\delta\rho(t) e^{i\omega_k t}\mathrm{d}t,
\end{align}
but the Fourier series using $\delta\hat{\rho}_k$ and $\delta\hat{\rho}^{(0)}_k$ are equal inside the simulation window.
The Fourier series \autoref{app:rho_fourier_series} is not, in general, zero outside the simulation time window $0 < t < T$.

Now, we consider a different pulse $v'(t)$ leading to a different response (new quantities denoted by primes)
\begin{equation}
    \delta\rho'(t) = \int_0^t \chi(t') v'(t-t')\mathrm{d}t',
\end{equation}
with Fourier coefficients defined
\begin{align}
    \hat{v}'^{(0)}_k &= \int_0^{T}v'(t) e^{i\omega_k t}\mathrm{d}t,
\end{align}
and, as before, the Fourier coefficients $\delta\hat\rho'_k = \hat{\chi}^{(0)}_k\hat{v}'^{(0)}_k$ reproduce $\delta\rho'(t)$ in the simulation time window $0 < t < T$.

Now we require that the Fourier coefficients of the new pulse are non-zero for every non-zero coefficient of the old pulse, so that we can write
\begin{align}
    v'^{(0)}_k = K^{(0)}_k v^{(0)}_k,
\end{align}
which is equivalent (according to the circular correlation theorem) to the pulses being related by the convolution
\begin{equation}
    v'^{(0)}(t) = \int_0^{T'} K^{(0)}(t')v^{(0)}(t-t')\mathrm{d}t',
\end{equation}
where $K^{(0)}(t)$ is some zero-padded function and $K^{(0)}_k$ its Fourier coefficients.
This means that the onset of the new pulse must be no later in time than the onset of the old pulse and introduce no new frequencies.

Then the Fourier series of the new response can be written
\begin{align}
    \delta\hat\rho'_k &= \hat{\chi}^{(0)}_kK^{(0)}_k\hat{v}^{(0)}_k
    = K^{(0)}_k \delta\hat\rho_k,
\end{align}
which is also equivalent to a convolution with $K^{(0)}$.
Because we are only interested in the response $\delta\rho'(t)$ during the simulation time $0 < t <  T$, and $K^{(0)}$ is zero-padded, we can thus swap $\delta{\hat{\rho}}_k$ for $\delta{\hat{\rho}}^{(0)}_k$ and calculate the new response as the Fourier series
\begin{align}
    \delta{\rho}'(t) = \frac{1}{T'}\sum_{k=-\infty}^\infty \delta{\hat{\rho}}^{(0)}_k \frac{\hat{v}'^{(0)}_k}{\hat{v}^{(0)}_k} e^{-i\omega_k t}.
\end{align}
We have used that $K_k = \hat{v}'^{(0)}_k / \hat{v}^{(0)}_k$ for every non-zero value of $\hat{v}'^{(0)}_k$.

In practice, we have sampled $\delta\rho(t)$ and $v(t)$ on a grid of $N$ times $t_j = j \Delta t$ with time step $\Delta t = T / N'$.
Then the Fourier coefficients are approximated as
\begin{align}
    \delta\hat{\rho}_k = \Delta t \left(\sum_{j=0}^{N-1} \delta \rho(t_j) e^{i\omega_k t_j}\right)
\end{align}
and
\begin{align}
    \delta\rho^{(0)}(t_j) = \frac{1}{\Delta t} \left(\frac{1}{N'}\sum_{k=0}^{N'-1} \delta\hat\rho_k e^{-i\omega_k t_j}\right),
\end{align}
where the terms inside the brackets are the discrete Fourier transform and inverse discrete Fourier transform respectively, and $N' \geq 2N$.
The approximation is good if the perturbation does not have any response above the Nyquist frequency $\omega = 2\pi / (2\Delta t)$; otherwise, aliasing effects are seen.

\end{document}